# Imaging cytometry without image reconstruction (ghost cytometry)

Sadao Ota *[a,b], Ryoichi Horisaki [c,d], Yoko Kawamura [b], Issei Sato [a,b,e], Hiroyuki Noji [a,d]

[a]University of Tokyo, 7-3-1 Hongo, Bunkyo-ku, Tokyo 113-8654, Japan
[b]ThinkCyte Inc., 7-3-1 Hongo, Bunkyo-ku, Tokyo 113-8654, Japan
[c] Department of Information and Physical Sciences, Graduate School of Information Science and Technology, Osaka University, 1-5 Yamadaoka, Suita, Osaka 565-0871, Japan.
[d]JST, 4-1-8 Honcho, Kawaguchi-shi, Saitama 332-0012, Japan.
[e]RIKEN AIP, 1-4-1 Nihonbashi, Chuo-ku, Tokyo, 103-0027, Japan.

## Abstract

Imaging and analysis of many single cells hold great potential in our understanding of heterogeneous and complex life systems and in enabling biomedical applications. We here introduce a recently realized image-free "imaging" cytometry technology, which we call ghost cytometry. While a compressive ghost imaging technique utilizing object's motion relative to a projected static light pattern allows recovery of their images, a key of this ghost cytometry is to achieve ultrafast cell classification by directly applying machine learning methods to the compressive imaging signals in a temporal domain. We show the applicability of our method in the analysis of flowing objects based on the reconstructed images as well as in that based on the imaging waveform without image production.

## I  Introduction

Imaging and analyzing a static number of single cells is of importance in a wide range of biological and medicinal research fields[1-3]. Despite remarkable advances in the developments of ultrafast imaging methods[4-7], speeding up imaging-activated cell sorting has been challenging mainly due to difficulties in compatibility of the acquisition of high dimensional image data and the image analysis including its reconstruction at the high speed in real-time.

We recently developed an image-free "imaging" cytometry approach, wherein we performed a direct analysis of compressive imaging waveforms with machine learning methods without image production, which we call ghost cytometry (GC)[8]. This GC skips the most computer-intensive process in the cell image analysis, which is reconstruction of two-dimensional images from raw signals, and thereby significantly relieved the computational bottleneck for the high-speed analysis and sorting based on morphological information[8]. Here we show the applicability of our method both in analysis of flowing objects based on the reconstructed images and in that based on the imaging waveform without image production. These results are compared with those obtained using a commercialized flow cytometer and an imaging flow cytometer.

## II  Body

### 1. Ghost motion imaging (GMI)

Compressive ghost imaging is a technique utilizing the compressive sensing methods to produce an image using a single pixel detector[9-11]. In our GMI, we adopted this ghost imaging technique to cells under motion: as cells are passing through a static, randomly patterned illumination area, fluorophores within the cells are continuously excited and the signals are detected by a single pixel detector. This measurement allows us to acquire temporally modulated waveforms of the fluorescence intensity from each cell experiencing multiple illumination spots at a time without switching the illumination pattern (**Fig 1A**). Lastly, combinatorial use of the acquired temporal waveform and a priori knowledge of the illumination pattern allows us to computationally reconstruct the cell image (**Fig 1B**).

### 2. Ghost cytometry (GC)

The more significance lies in our demonstration wherein the morphology-based cell classification was realized by directly applying machine learning-based analysis to the temporal waveforms without producing cell images (**Fig 1C**)[8]. GC starts with training of a machine learning-based classifier: many temporal waveforms of ghost imaging signals were acquired and labeled by cell types, fluorescence or other markers, to be used as a training data set. In testing, in turns, the trained model predicts the label from the imaging signals without using information of the labels. We used support vector machine (SVM) in this work.

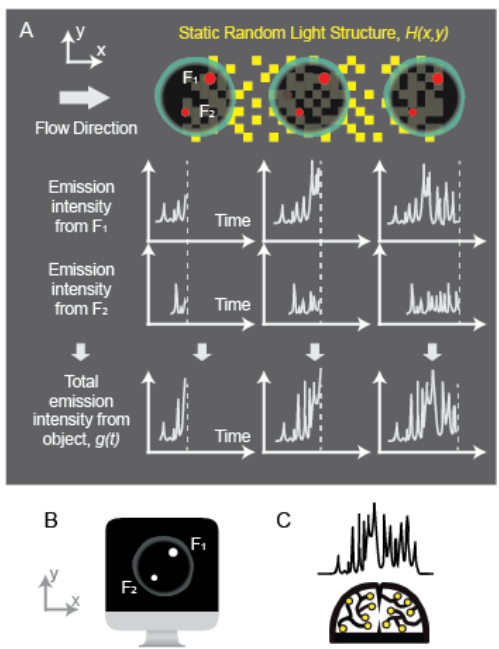

Figure 1. (A) Compressive ghost imaging process of objects under motion, wherein total fluorescence emission intensity is temporally modulated and recorded by a single pixel detector as the flowing cell experiences different excitation patterns according with time. While the image can be recovered (B), by training and testing a machine learning model, the imaging waveform signals can be directly analyzed and classified without image recovery (C).

### III  Results and Discussion

**1. GMI-based image recovery of flowing fluorescent beads and its analysis**

We show image-production capability of GMI by acquiring and analyzing images of flowing fluorescence beads having different size and intensity (**Fig 2**). We performed the GMI of 600 fluorescent beads, consisting of 300 each of two different sizes and intensities, under a hydrodynamic focus in flow. Using an experimentally measured illumination pattern, we computationally processed the waveforms to build a library of the GMI images (**Fig 2A**). From this library, we analyzed singlet bead images to determine bead diameter and total fluorescence intensity by Canny edge detection and integrating the cropped images, respectively. As shown in **Fig 2B**, obtained histograms of both diameter and fluorescence intensity produce a characteristic bimodal distribution with peaks at the expected diameters and intensities. Furthermore, when we compare these histograms with those of forward scatter and total fluorescence intensity measurements of the same bead samples using a standard flow cytometer (JSAN, Bay Bioscience), gated to include only singlets, the results show good consistency (**Fig 2C**).

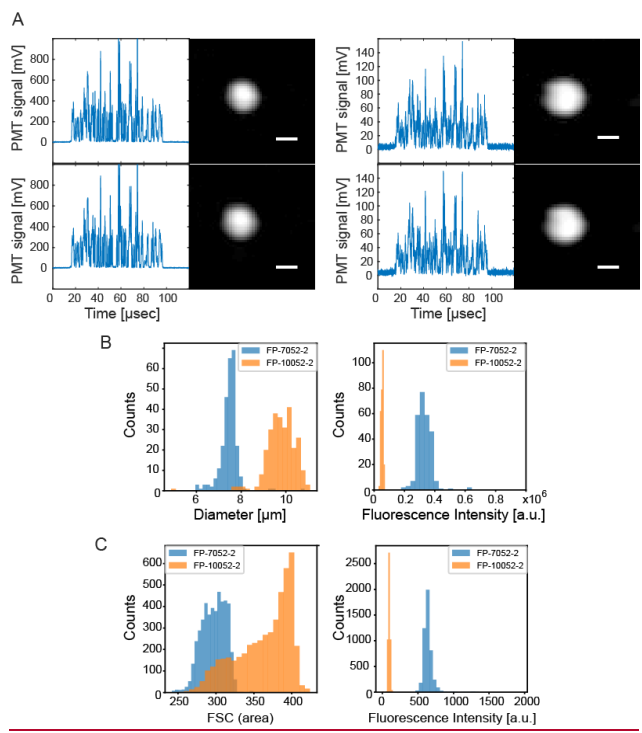

Figure 2. GMI image reconstruction and analysis in flow implementations. (A) Example waveforms from fluorescence beads in a high-speed flow and their reconstructed images (cropped). Left and right panels were from different kinds of beads (Fluorescent Yellow Particles, FP-7052-2 and FP-10052-2, Spherotech, with mean diameters of 7.4 μm and 10.2

μm, respectively). (B) Left and right panels are histograms of diameters and total fluorescence intensity, respectively. We obtained them by using the single-bead images from a library of the reconstructed images. We cropped the reconstructed images into a size of 32 by 32 pixels and calculated their diameters and total intensities by Canny edge detection and integrating the cropped images, respectively. (C) Left and right panels are histogram of forward scattering and fluorescence intensity, respectively, measured by using a commercialized flow cytometer (JSAN). All scale bars = 5 μm.

Moreover, we performed PSNR measurements of fluorescence beads in the PDMS device used previously[8]. Using a reference bead image taken by a conventional microscope (**Fig 3A**), we obtained moderately good PSNRs of 28.8 dB and 29.0 dB for two example reconstructed bead images, respectively (**Fig. 3B**).

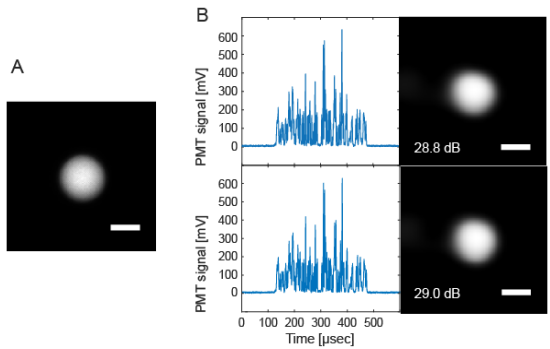

Figure 3. (A) Left panel is a fluorescence image of a fluorescent bead (FP-7052-2), taken by a conventional microscope (Olympus), used as a reference for calculating PSNRs. This image was taken on a slide glass with a 20x objective (UPLSAPO 20x, Olympus) and an arrayed pixel camera (ZWO ASI1600MM). (B) Right panels show example compressive waveforms from the same kind of beads flowing in a PDMS fluidic device and their reconstructed and cropped images. The PSNRs were calculated with images adjusted to scale, cropped to 134 by 134 pixels, and normalized by intensity. Each bead image shown here is normalized with their max intensities for enhancing their visibilities. All scale bars = 5 μm.

While achieving moderately good PSNRs, it is also worth noting that, in general, image recovery in compressive sensing is computationally costly, and image qualities are affected by calibration errors rather than the theoretical limit. Our image-free imaging cytometry bypasses both of these issues by utilizing machine learning directly on the waveforms.

## 2. GC-based image-free classification of cells

Using the training and testing process of GC, which is concisely described above, we classified MCF-7 and MIA PaCa-2 cells that are morphologically similar but different. In the experiment, both cells were fluorescently stained using green dyes (LIVE/DEAD™ Fixable Green Dead Cell Stain Kit, for 488 nm excitation), measured at short temporal width of < 100 μsec, and classified by applying the machine learning model on the fluorescence "imaging" waveforms. The classification result recorded the area under a receiver operating characteristic (ROC) curve (AUC) of 0.971 over about fifty thousands of cells[8].

In addition, we separately confirmed the morphological characteristics of the MCF-7 and MIA PaCa-2 cells to further understand the high accuracy of our GC-based classification. It is well known that different samples of even the same cell lines can have different sizes[12-13], depending on various conditions including culture and fixing methods. In our case, the two cell lines are experimentally shown similar in average size (**Fig 4A**) and have different morphological characteristics (**Fig 4B**). In the experiment, the cells were fixed and stained using the green dyes of fixable green. By flowing the two cell types separately into a commercialized flow cytometer (JSAN), we first measured the peak intensity of the green fluorescence and the forward scattering signals from each cell. The resultant plots in **Fig 4A** shows that the two cell types share their fluorescence intensity and size. By flowing the two cell types separately into an image flow cytometer (ImageStreamX, 20x objective lens), we then generated a library of fluorescence images of the cell populations. We preprocessed the data set in this image library using IDEAS® software (Merck Millipore Inc.) to identify a population of focused single cells[14] before cropping raw images around the center of mass into images of 28 x 28 pixels images and randomly rotating them with a multiple of 90 degrees. 1,000 of single MCF-7 cells and 1,000 of single MIA PaCa-2 cells were used for training, respectively, while 100 of single MCF-7 cells and 100 of single MIA PaCa-2 cells were used for testing, respectively. To the image dataset which was lastly normalized in a cell-type independent way, we performed an image-based SVM classification. The AUC obtained was 0.967, confirming that the cell types are morphologically distinguishable even at relatively low (20x) magnification, corresponding to an image pixel size of approximately 1μm.

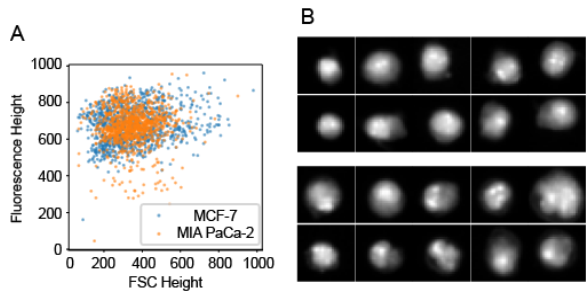

Figure 4. (**A**) Scattering plot of the intensity height of green fluorescence and the height of forward scattering intensity for MCF-7 and MIA PaCa-2 cells, obtained using a commercialized flow cytometer (JSAN), showing that the cell lines are similar in the fluorescence intensity and size. (**B**) Example images of the data set (top ten images are MCF-7 cells and bottom ten images are MIA PaCa-2 cells), obtained using a commercialized image flow cytometer (ImageStreamX, using 20x objective lens). This image data set was used for an image-based classification using the SVM-based model.

Combining the comprehensive discussions above, the two cell lines are similar in average size but have different morphological characteristics. In addition to this morphological difference, the classifier could detect other factors such as flowing velocities, which can be encoded in the waveforms as well. Such velocity difference could arise from morphological characteristics such as size and shape, as well as other physical characteristics such as deformability. The effect of such velocity difference also may remain in the reconstructed images in other reported flow-scanned imaging flow cytometry technologies unless the obtained imaging signals are appropriately normalized. Direct learning of waveforms without image production is not degraded by the velocity difference; the image-free approach utilizes such difference for a more robust classification. In addition to further exploring the potential of image-free approach for extracting essential cell features, we thus think that analysis of the whole waveform holds potential to utilize various types of characteristics of cells comprehensively.